\documentclass[10pt, conference, compsocconf, redeflists]{IEEEtran}

\usepackage{booktabs} 
\usepackage{color}
\usepackage{subcaption}
\usepackage{breqn}
\usepackage{graphicx}
\usepackage{tabularx}
\usepackage{algorithm}
\usepackage{algorithmicx}
\usepackage{algpseudocode}
\usepackage{etoolbox}\AtBeginEnvironment{algorithmic}{\small}

\let\oldReturn\Return
\renewcommand{\Return}{\State\oldReturn}

\let\OLDthebibliography\thebibliography
\renewcommand\thebibliography[1]{
  \OLDthebibliography{#1}
  \setlength{\parskip}{2pt}
  \setlength{\itemsep}{2pt plus 0.3ex}
}

\begin{document}
\title{Data Context Informed Data Wrangling}

\author{\IEEEauthorblockN{Martin Koehler\IEEEauthorrefmark{1},
		Alex Bogatu\IEEEauthorrefmark{1},
        Cristina Civili\IEEEauthorrefmark{2},
		Nikolaos Konstantinou\IEEEauthorrefmark{1},
		Edward Abel\IEEEauthorrefmark{1},
        Alvaro A. A. Fernandes\IEEEauthorrefmark{1},}
        \IEEEauthorblockN{
		John Keane\IEEEauthorrefmark{1},
        Leonid Libkin\IEEEauthorrefmark{2} and
        Norman W. Paton\IEEEauthorrefmark{1}}
        \IEEEauthorblockA{\IEEEauthorrefmark{1}School of Computer Science, University of Manchester, Manchester, UK}
		\IEEEauthorblockA{\IEEEauthorrefmark{2}School of Informatics, University of Edinburgh, Edinburgh, UK}    
}

\maketitle

\begin{abstract}
The process of preparing potentially large and complex data sets for further analysis or manual examination is often called data wrangling. In classical warehousing environments, the steps in such a process have been carried out using Extract-Transform-Load platforms, with significant manual involvement in specifying, configuring or tuning many of them. 
Cost-effective data wrangling processes need to ensure that data wrangling steps benefit from automation wherever possible. 
In this paper, we define a methodology to fully automate an end-to-end data wrangling process incorporating data context,  which associates portions of a target schema with potentially spurious extensional data of types that are commonly available. Instance-based evidence together with data profiling paves the way to inform automation in several steps within the wrangling process, specifically, matching, mapping validation, value format transformation, and data repair. The approach is evaluated with real estate data showing substantial improvements in the results of automated wrangling.
\end{abstract}


\IEEEpeerreviewmaketitle

\section{Introduction}
\label{sec:introduction}

Data wrangling is the process by which potentially large and complex data sets are prepared for analysis or manual examination~\cite{Furche2016,Kandel2011}. However, there may be quite a few steps involved in data wrangling; a possible process involves data extraction (for example from the deep web or web tables), schema matching, mapping generation, data repair, value format transformations, and resolution and fusion of entities.

Such steps can be carried out using Extract-Transform-Load (ETL) \cite{Vassiliadis2009} or Big Data analytics platforms \cite{Kaniovskiy2017}, both necessitating significant manual involvement in specifying, configuring, programming or tuning many of the steps.  It is widely reported that intense manual involvement in such processes is expensive (e.g. \cite{Kandel2011}), often representing more than half the time of data scientists. As the numbers of data sources within organisations and in the public domain grows, there is an increasingly pressing need for cost-effective techniques for addressing the variety and veracity of big data.

In this paper, we study the problem of automating an end-to-end data wrangling process, that is, to integrate (addressing variety) and clean (addressing veracity) a large set of input sources and create a data product that is suitable for downstream analysis. In more detail, we focus on how a wrangling process can be improved by {\it data context}: data from the domain in which wrangling is taking place~\cite{Furche2016}. There have been proposals tailored to a specific type of auxiliary information and for automating individual steps in the wrangling process 
(e.g. \cite{Furche2014, Abedjan2016DataXFormer, Aumueller2005}), 
but there is a need to be more systematic and holistic, ensuring that 
all the steps can be automated, and that all these steps make use of the available data context. 

Automating the end-to-end process supports the population of the data product, but potentially results in a data product of limited quality. Data context, such as master data, reference data, or example entities from the domain of data wrangling, can serve as a guide to improve the results of many steps within the wrangling process. Specifically, the claim is that a small number of often readily available types of contextual data can substantially improve the quality of the automatically produced data product. 

Our solution adopts and extends some of the latest techniques from the data profiling, integration and cleaning communities on dependency discovery \cite{Rostin2009,Papenbrock2016,Fan2011}, instance-based schema matching \cite{Aumueller2005}, mapping generation and validation \cite{Bonifati2008}, value format transformations \cite{Bogatu2017}, and rule-based repair \cite{Cong2007}. We refine and combine their approaches to the use of target instances for automation to provide a comprehensive, end-to-end approach incorporating instance based evidence from the \textit{data context }that may be partial.

Our contributions in this paper are as follows:
\begin{enumerate}
	\item A definition of the notion of {\it data context}, and of its specific types. 
	\item A methodology to fully automate an end-to-end data wrangling process that incorporates data context.
	\item A description of how the data context can inform multiple steps within an end-to-end wrangling process, specifically matching, mapping validation, value format transformation, and rule-based data cleaning.
	\item An evaluation of the approach in a real estate application combining deep web and open government data that shows both: (i) significant improvements in the results of automated processes (e.g., the precision of the result increased from {\it 0.64} to {\it 0.88}); and (ii) the impact of data context on the individual steps. 
\end{enumerate}

\section{Problem statement} 
\label{sec:problem}

\begin{figure}[t]
	\centering
	\includegraphics[width=0.98\columnwidth]{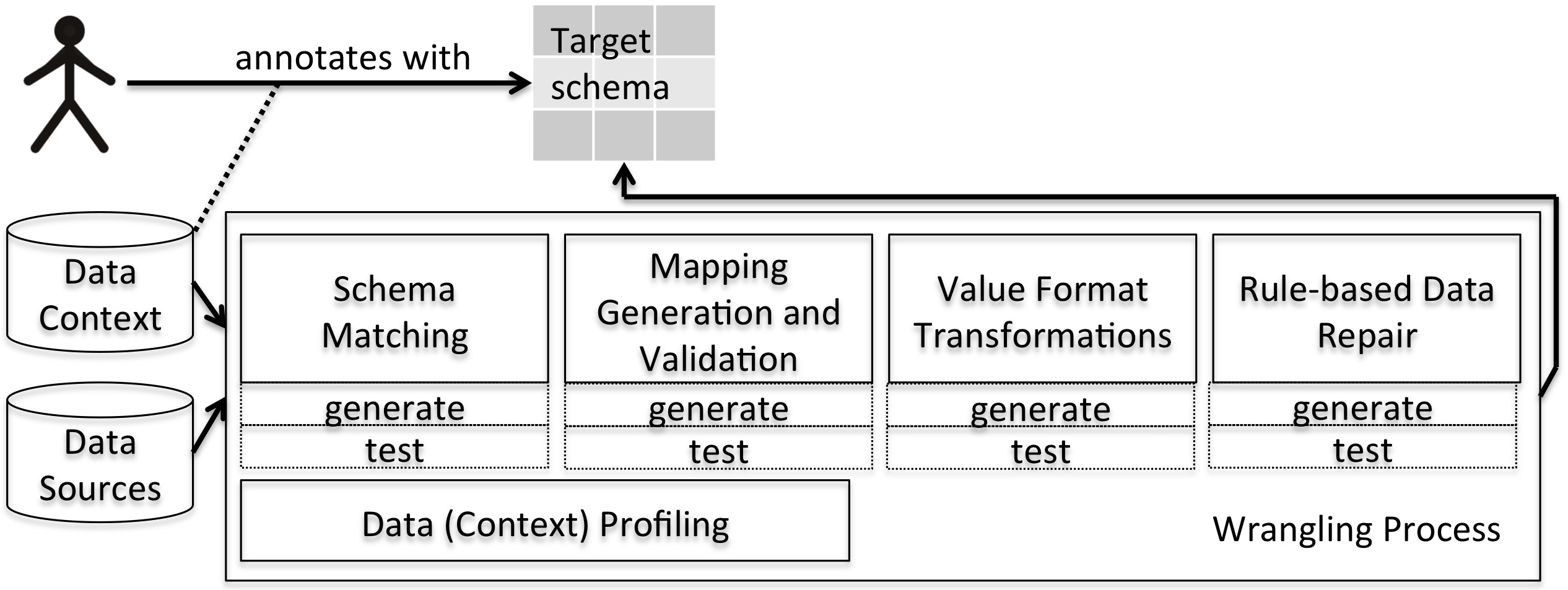}
	\caption{Data wrangling process: User annotates the target schema with data context and starts the wrangling process that involves several steps.}
	\label{fig:arch}
\end{figure}

Although data wrangling processes may include different steps, in this paper we demonstrate the application of data context\footnote{Note that our notion of {\it data context} is different from, and complementary to, previous work on context aware systems (e.g. \cite{Bolchini2009}).  In such proposals, the focus is on identifying the subset of an extent that is most appropriate for a user in a given situation. In contrast, our notion of {\it data context} emphasizes data from the domain within which data wrangling occurs. The term {\it data context} is also used in the proposal for the Ground {\it data context service} \cite{Hellerstein-17}, which is used to capture metadata and annotations relating to diverse data sets. Our notion of data context would seem to be suitable for capturing and sharing using a platform such as Ground.} using the data wrangling process illustrated in Figure \ref{fig:arch}. We assume that the end user is a data scientist, who is familiar with the domain within which the data is to be wrangled, and thus who can provide a data product schema $(P)$ that is to be populated by the wrangling process. An example target schema for the real estate domain is illustrated in Table \ref{tab:transformdata}. Given some data sources $S$, such as those in Table \ref{tab:rawdata}, it is the role of the wrangling process to populate the target schema with (as much as possible) correct and consistently formatted values.  

In the process in Figure \ref{fig:arch}, this involves matching the source and target schemas, generating mappings from the matches, reformatting values that may be represented in different ways (e.g. transforming {\it Homestead Rd} to {\it Homestead Road}) and completing or correcting inconsistent values (e.g. providing a missing {\it city} value of {\it London} for a property with a {\it London} {\it postcode}). Carrying out all these tasks automatically is not straightforward. 

In this paper, we describe how automation can be informed by the data context consisting of data sources $(D)$ that can be aligned with the target schema, thereby providing partial, potentially erroneous instance-based evidence about the target. Data context data (examples in Table \ref{tab:datacontext}) can be: 

\setlength{\belowcaptionskip}{-10pt}
\small

\begin{table*}[tb]
	\footnotesize
	\begin{tabular}{|l|l|l|l|l|l|}
		\cline{1-1}
		Zoopla \\ 
		\hline		heading\_h1&h2\_nth\_of\_type\_1&n\_th\_of\_type&h2\_nth\_of\_type\_1&details\_box\_8&details\_box\_6\\
		\hline 
		Whitfield St & Greater London & W1T 5EF &137,495&Leaders&898756\\
		Biscayne Ave & London & E14 9BE&189,950&ReedsRains&8654789\\
		\hline 
		Belvoir \\
		\hline
		lst\_det\_address\_h2&lst\_det\_city\_h2&lst\_details\_h1&p\_nth\_of\_type&tab\_details\_ui\_tabs\_pc&tab\_details\_ui\_tabs\\
		\hline
		9 Canton St & London & E14 6JW &595,000&E14 3NE&Limited Belvoir London\\
		20 South Drive & - & W1A 0AA &575,000&E14 3NE&Limited Belvoir London\\
		\hline
		Deprivation \\
		\cline{1-4}
		postcode&postcodestatus&crimerank&crimedecile \\
		\cline{1-4}
		E14 3NE&Live&34&7\\
		\cline{1-4}
	\end{tabular}
	\caption{A collection of raw web extracted real-estate records and deprivation data}
	\label{tab:rawdata}
	\normalsize
\end{table*}
\setlength{\belowcaptionskip}{0pt}
\normalsize

\textit{Reference data:}
A collection of values that stipulate the valid domain of a set of specific attributes of the product $P$: correctly repaired and transformed instances $I_P$ of $P$ are a subset of instances $I_d$ of $d \in D$, for the set of related attributes. Thus reference data is complete, in that there are no missing values, and accurate, in that it provides correct values to be occurring in the product. In the real estate example in Table \ref{tab:datacontext}, the {\it Address data} is used as reference data, and is freely available in the UK\footnote{Open addresses: https://alpha.openaddressesuk.org/}. 

\textit{Master data:}
Master data can be defined as constituting a consistent view on the core entities in an organization. Thus, master data are correct and accurate values stipulating a set of target attributes. In contrast to reference data, the set relation between the sets $I_P$ of $P$ and $I_d$ of $d \in D$ is not known \textit{a priori}. 
In the real estate example, master data comprises 
information about the properties for sale/rent from the real-estate agency the data scientist works for.

\textit{Example data:} 
A collection of data items that (partly) express the domain of the target. Examples may include empty and erroneous values and stipulate a set of target attributes. Again, the relation between the sets $I_P$ and $I_d$ is not known \textit{a priori}. In the real estate example in Table \ref{tab:datacontext}, the freely available UK price paid data\footnote{Price paid: https://www.gov.uk/government/collections/price-paid-data} holds information about sample property sales conducted in the past and thus exemplifies values for attributes of interest. 

Data context sources $(D)$ can be related to the tables in the target schema $(P)$ using {\it data context relationships} $R(d,p,t)$, where $d \in D$, $p \in P$ and $t \in \{reference, master, example\}$. For expressing data context relationships we use the notion of a \textit{tuple generating dependency} ({\it tgd}) of the form $\forall x(\phi_D(x) \to \exists y \psi_P(x,y))$ where $\phi_D(x)$ is a conjunction of atoms over the data context and $\psi_P(x,y)$ is a conjunction of atoms over the target schema. As an example, considering data context sources $(D)$ in Table \ref{tab:datacontext} and the data product $(P)$ in Table \ref{tab:transformdata}, {\it Address data} (A), as reference data, can be used to specify the valid domain of attributes in a target table (P), by means of the tgd $\forall p, sn, tn, pn (A(p, sn, tn, pn) \to \exists y_1,y_2,y_3,y_4 P (sn, tn, pn, y_1, y_2, y_3, y_4))$, giving rise to a data context relationship $R(A, P, reference)$. The data product is not directly populated from the data context, but rather the data context is used to inform the steps that populate the data product. We assume the data scientist has sufficient knowledge of the domain to identify suitable data sets for the data context, and to envisage their precise relationship to the target schema (as exemplified by the tgd above). They might then be made available using an user interface to the wrangling platform, as done in VADA \cite{Konstantinou2017}.

We can now explicitly state the problem of how to inform an end-to-end data wrangling process consisting of multiple steps with data context:
given a set of sources $S$, source instances $I_s$ for each source $s \in S$, a target $P$, data context sources $D$ with instances $I_d$ when $d \in D$, and data context relationships $R(d,p,t)$, automatically populate the target $P$ with (potentially transformed and repaired) instances. 

\section{Data Context informed Wrangling} 
\label{sec:practice}

This section describes how data context is used to inform the automation of the stages in the wrangling process (see Figure \ref{fig:arch}). For each consecutively executed stage, we define the problem and present a general methodology with two phases: \textit{generation} -- the creation of candidate solutions; \textit{testing} --  assessment, selection and refinement of candidates in the light of the evidence from data context. We show how each stage can be automated without data context, and how the approach can be revised to take account of data context. 

\subsection{Schema Matching}
\label{subsec:prac:matching}

\setlength{\belowcaptionskip}{-10pt}
\begin{table}[tb]
	\footnotesize
	\begin{tabular}{|l|l|l|l|l|l|l|}
		\cline{1-2}
		\multicolumn{2}{|c|}{Price paid data} \\
		\hline
		price\_paid & saon & paon & street & postcode & town\\
		\hline
		155000 & Flat 6 & 25 & Bournem. Rd & SE15 4UJ & London\\
		\hline
	\end{tabular}

	\begin{tabular}{|l|l|l|l|l|l|}
		\cline{1-2}
		\multicolumn{2}{|l|}{Address data} \\
		\hline
		pao & street.name & town.name & postcode.name\\
		\hline
		14 & Heron Lane & Scarborough & YO12 4TW\\
		\hline
	\end{tabular}  
    \quad
    \begin{tabular}{|l|l|l|l|l|l|l|}
		\cline{1-1}
		Master data \\
		\hline
		street\_nr & paon & city & postcode & price\\
		\hline
		Redhill street & 8 & London & E14 3NE & 125.000£\\
		\hline
	\end{tabular}
	\caption{Data context information: price paid data (examples), address data (reference data) and master data}
	\label{tab:datacontext}
\end{table}
\normalsize
\setlength{\belowcaptionskip}{0pt}

\textit{Problem definition.} Schema matching can be defined as the problem of detecting schematic correspondences between schema elements of data sources $S$ and the target $P$. Schematic correspondences identify potentially equivalent pairs of schema elements, along with a confidence measure that is most often expressed as a similarity score. 

\textit{Evidence.} We approach the challenge of automating schema matching by applying the generate-and-test methodology. Generating and testing candidate schematic correspondences involves different types of evidence. \textit{Metadata evidence} explains characteristics of schema elements such as their names, data types, and structural properties, and supports the comparison of the source and the target schema for finding correspondences. \textit{Target instances} provide additional evidence on the values that are part of the target, which can be exploited by instance-based matchers. \textit{Domain-specific evidence} explains additive knowledge of parts of a data source. Usually, domain evidence is created and maintained by domain experts and exploited by domain recognisers or gazetteers. 

\textit{Context informed automation.} Algorithm \ref{algo:match} is used to automate schema matching, using data context information when it is available. The algorithm is invoked for each source $S$ and the target $P$. It uses metadata and data context based evidence in the two phases, generate, and test. 

In the absence of data context, the algorithm applies schema based matchers (line 2) to generate candidate schematic correspondences. We utilize the Coma 3.0 community edition\footnote{Coma Community Edition: https://sourceforge.net/projects/coma-ce/}, specifically, the Coma workflow (configuration $7001$) combining different metadata-based match heuristic. When data context is provided in $D$, each such data set is used as a partial extensional representation of the target to carry out instance based matching with the source (line 5). Specifically, the Coma instance matchers (configuration $7008$) are executed in addition to the schema-based matchers. 

Match testing takes further advantage of data context through the utilization of domain recognisers. In general, recognisers employ dictionaries or rules to recognise the data values of certain kinds of attributes. In our system, we have implemented generic recognisers, exploiting the information gained from data context (line 10). The generic recognisers combine inference of basic types (e.g. numeric, floating-point, string) and characteristics such as the length and tokenization of the values. Recognisers are utilised to refine schematic correspondences to target attributes aligned with data context by increasing or decreasing their similarity scores, and to detect new correspondences not detected by schema- and instance based matchers (line 11). 

\setlength{\belowcaptionskip}{-10pt}
\footnotesize
\begin{center}
\begin{table}[t]
	\begin{tabular}{|m{0.6cm}|m{2cm}|m{1cm}|m{1.5cm}|m{1cm}|}
		\hline
		tuple&street&city&postcode&price\\
		\hline
		$t_1$&Whitfield Street & London & W1T 5EF &137,495\\
		$t_2$&Biscayne Ave & London & E14 9BE&189,950\\
		$t_3$&Canton Street & London & E14 6JW&595,000\\
		$t_4$&South Drive & London & W1A 0AA&575,000\\
		\hline
	\end{tabular}
	\begin{tabular}{|m{0.6cm}|m{3cm}|m{1.5cm}|m{1.45cm}|}
		\hline
		tuple& agency & contact & crimestats\\
		\hline
		$t_1$&Leaders&898756 & 136\\
		$t_2$&ReedsRains&8654789 & 45\\
		$t_3$&Belvoir London LTD&E14 3NE& 34\\
		$t_4$&Belvoir London LTD&E14 3NE& 78\\
		\hline
	\end{tabular}
	\caption{Transformed, integrated and repaired records}
	\label{tab:transformdata}
	\normalsize
\end{table}
\end{center}
\normalsize
\setlength{\belowcaptionskip}{0pt}

\begin{algorithm}[b]
	\caption{Data context matching}\label{algo:match}
	\begin{algorithmic}[1]
		\Require source schema $S$ and instances $I_s$, product schema $P$, Set of data context schemas $D$ and instances $I_D$, lower and upper bound $lb, up$
		\Ensure Set of matches $M$
		\Procedure{match}{}
		\State $M \gets gen\_schema\_match(S,P)$
        \State $M_d \gets M_r \gets \{\}$
		\ForAll {$d \in D$}
			\State $I_P \gets I_d$
			\State $M_d \gets combine(M_d,gen\_instance\_match(S,P))$
		\EndFor
		\State $M \gets update(M, M_d)$
		\ForAll {$d \in D$}
			\State $M_r \gets match\_domain(S,P,d)$
			\State $M \gets test(M, M_r, lb, up)$
		\EndFor
		\Return $M$
		\EndProcedure
	\end{algorithmic}
\end{algorithm}

\subsection{Schema Mapping Generation and Validation}
\label{subsec:prac:mapgen}

\textit{Problem definition}
Schema mapping generation and validation can be defined as the problem of generating data transformations from data sources $S$ into a target $P$ and validating the resulting candidates for use. Schema mappings can be expressed using source-to-target tuple generating dependencies (st-tgds) of the form 
$\sigma: \forall x(\phi_S(x)$ $\to \exists y \psi_P (x,y))$, where $\phi_S(x)$ 
is a conjunction of source atoms, and $\psi_P (x,y)$ is a conjunction of target atoms.

\textit{Evidence.} We approach the challenge of automating schema mapping generation and validation by applying generate and test phases. Generating and testing are based on different types of evidence. \textit{Metadata evidence} describing schema elements, their structure and primary/foreign key relationships, combined with schematic correspondences between the sources and the target,  supports the application of mapping generation approaches such as Clio~\cite{Miller2000} or ++Spicy~\cite{Marnette2011}. \textit{Data profiling}~\cite{Rostin2009,Papenbrock2016} infers descriptive information about sources that can be exploited by mapping generation and validation, though automatically detected candidate keys and inclusion dependencies can provide misleading evidence. 
\textit{Target instances} can be exploited by mapping validation approaches exploiting instance based similarity measures. For example, tree similarity measures taking into account the topology and the information content support target instance evidence~\cite{Bonifati2008}.

\textit{Context informed automation.}
Algorithm \ref{algo:mapping} is used to automate mapping generation and validation without data context, and improves the result if data context is available. The algorithm is invoked for the set of input sources $S$ and the target. It takes metadata, profiling and data context evidence into account. In the generation phase (line 2), foreign key candidates are detected based on criteria from \cite{Rostin2009}. Foreign key candidates are generated if we identify a key candidate and an inclusion dependency between two attributes in different sources matching with the same target attribute. 

The input sources are clustered into subsets exploiting the candidate foreign keys (line 3), and mapping generation and validation is executed for each source cluster (line 9). To generate candidate mappings we use the ++Spicy toolkit \cite{Marnette2011}, which generates mapping candidates based on the given schematic correspondences and foreign key candidates (see Section \ref{subsec:prac:matching}). We generate mappings by exploiting and neglecting the detected foreign keys to handle potentially spurious evidence.

An example candidate mapping generated for the data sources \textit{Zoopla (Z)} and \textit{Deprivation (D)} in Table \ref{tab:rawdata} and the data product in Table \ref{tab:transformdata} is depicted in (\ref{eq:sttgd}).

 \begin{dmath}\label{eq:sttgd}
	\sigma: \forall x_{1,..,9}( Z(x_1, x_2, x_3, x_4,x_5,x_6) 
	\land D(x_3, x_7, x_8,x_9), \to \exists y_1,y_2 P(y_1, x_1, x_2, x_3, x_4, x_5, y_2, x_8))
 \end{dmath}

When data context is available (line 10), we test mapping candidates by carrying out mapping verification, as proposed in \cite{Bonifati2008} for target instances, to select between alternative candidates. For each candidate mapping we translate the data context instances into the target and compute the verification score by executing structural analysis. 

The algorithm repeats the generate and test phases, lowering the threshold for schematic correspondences (line 7), retaining the result with the best verification score (line 13). In the absence of data context, we choose the mapping candidate satisfying the most schematic correspondences for each group of sources.

\begin{algorithm}[t]
	\caption{Mapping generation and validation}\label{algo:mapping}
	\begin{algorithmic}[1]
		\Require set of source schemas ${S}$ and instances ${I_S}$, product schema $P$, Set of data context schemas $D$ and instances $I_D$, set of relationships $R$, upper bound $up$, lower bound $lb$, step size $step$, set of matches $M$
		\Ensure Set of mappings $G$
		\Procedure{mapping}{}
		\State $K \gets profile\_fks({S})$
		\State $C \gets cluster(S,K)$
		\State $G_{best} \gets \{\}$
        \State $t \gets ub$
		\ForAll {$c \in C$}
			\While{$t > lb, t=t-step$}
				\State $M_t \gets filter(M,t, m.score > t)$
				\State $G \gets gen\_schema\_mapping(c, M_t,K)$
				\ForAll {$d \in D$}
					\State ${V_g} \gets test({G},{D},{R},{I_D})$
				\EndFor
				\If {$max({V_g} > G_{best_{c}})$}
					\State $G_{best} \gets max({V_g})$	
				\EndIf
			\EndWhile
		\EndFor
		\Return $G_{best}$
		\EndProcedure
	\end{algorithmic}
\end{algorithm}

\subsection{Value Format Transformation}
\label{subsec:prac:transform}

\textit{Problem definition.} The value format transformation problem is that of converting the values for an attribute from sources into a uniform format in the target by applying syntactic manipulations such as \textit{concatenate} or \textit{substring}. For example, a source might abbreviate recurring parts of an address (e.g. {\it Canton St}), when the full representation is required ({\it Canton Street)}.  Correctly inferring and applying transformation rules is a hard problem and usually involves some form of user involvement~\cite{Gulwani2011,Kandel2011}. 
In our approach, we seek to automatically identify data examples that can be used to synthesize transformation rules using FlashFill~\cite{Gulwani2011}.

\textit{Evidence.} We approach the challenge of automating value format transformations by applying the generate and test methodology. Generating and testing are based on different types of evidence. \textit{Metadata evidence} describing schema elements combined with schematic correspondences between the source and the target are used to identify potentially equivalent concepts whose value representations have to be aligned.
\textit{Profiling evidence} such as functional dependencies, gives rise to a means of aligning tuples from multiple data sets~\cite{Bogatu2017}. In this paper, data context provides the extensional data that is required for the target. More specifically,
assume that we would like to transform the values in source attribute $S.s_n$ into the format used in $P.p_m$. If functional dependencies (FD) in $S$ and in $P$ ($fd_1: [S.s_i] \rightarrow [S.s_n]$ and $fd_2: [P.p_j] \rightarrow [P.p_m]$) and schematic correspondences exist between the determinants 
($S.s_i, P.p_j$) and the dependents ($S.s_n, P.p_m$) of the functional dependencies, and if the values in the determinants correspond, data examples can be generated automatically. 

\textit{Context informed automation.} Algorithm \ref{algo:transform} is used to automate value format transformations when data context is available. To automate value format transformation, we make use of data context as a partial representation of the target.  
The algorithm is invoked for each source $S$ separately and returns a transformed source $S'$. In the generation phase, data examples $E$ between the source and each data context source $D$ are generated (line 6). Data examples are then utilized to automatically generate transformation rules (line 7). The algorithm presented in \cite{Gulwani2011,Bo2016} is used to synthesize transformation rules, from the identified source and target data examples, by applying a programming-by-example (PBE) approach. 

Data examples and transformation rules are tested and selected together using a k-fold validation approach. Different data context types are utilised consecutively to generate data examples and transformation rules. If source columns can be transformed by means of multiple data context items, we select the one with the larger set of data examples.  Finally, each selected transformation rule is applied on the sources to transform the data from the complete column.

\begin{algorithm}[t]
	\caption{Value format transformation}\label{algo:transform}
	\begin{algorithmic}[1]
		\Require source schema $S$ and instances $I_S$, Set of data context schemas $D$ and instances $I_D$, set of matches $M$
		\Ensure Transformed source $S'$
		\Procedure{transform}{}
		\State $FD_S \gets profile\_functional\_dependencies(S)$
		\State $R \gets \{\}$ \Comment{$R$ are selected transformation rules}
		\ForAll {$d \in D$}
			\State $FD_d \gets profile\_functional\_dependencies(d)$
			\State $E \gets generate\_examples(s,d,M,FD_S, FD_d)$
			\State $R_d \gets generate\_transform\_rules(E)$
			\State $R_d \gets test\_examples\_rules(E, R_d)$
			\State $R \gets accumulate\_selected(R_d)$
		\EndFor
		\State $S' \gets apply\_transform(S, R)$
		\Return $S'$
		\EndProcedure
	\end{algorithmic}
\end{algorithm}

\subsection{Rule Based Data Repair}
\label{subsec:prac:repair}

\textit{Problem definition} The data repair problem involves detecting and repairing certain classes of data errors, e.g. violations of integrity constraints. Integrity constraints can be given in a range of languages, varying from user-defined functions to database constraints like conditional functional or inclusion dependencies \cite{Abedjan2016,Ganapathi2016}. Here we adopt conditional functional dependencies (CFD). A CFD $\phi = (R: X \rightarrow Y, t_p)$ extends standard functional dependencies (FDs) by enforcing pattern tuples $t_p$ of semantically related constants~\cite{Cong2007}. To increase the consistency and accuracy  of data, violations have to be detected based on the given constraints, and suitable repairs have to be chosen for the detected violations. 

\textit{Evidence.} We approach the challenge of automating rule based data repair by applying generate and test phases. \textit{Target instances} can be used to underpin the automatic discovery of integrity constraints from data. For instance, the CFD discovery algorithm described in \cite{Fan2011} is capable of finding a canonical cover of s-frequent minimal constant CFDs based on an input relation $R$ and a support size $support$. The support size is the number of tuples matching the pattern of each CFD learned by the algorithm.

\textit{Context informed automation.} Algorithm \ref{algo:repair} is used to automate rule-based repair when data context is available. The algorithm is invoked for each source $S$ and the data product $P$. It takes source instances $I_S$ and data context into account. In the candidate generation phase, we generalize an approach that utilizes master data to discover certain fixes \cite{Fan2010} towards discovering CFDs for all available data context sources $D$. This relaxes the notion of certain fixes, as data context might provide spurious evidence. To fully automate the process, there is a need to (1) automatically configure the CFD discovery algorithm, and to (2) select the CFDs to apply. The described algorithm focuses on discovering and selecting a set of CFDs maximizing the precision of the repair process, i.e. minimising the number of incorrect repairs. We incrementally increase the support size parameter used for discovering CFDs (line 6, 7, 13) and apply filter and validation steps (line 7, 8) on the discovered ones. CFDs violating tuples in the training data are filtered out. The confidence of the CFDs is used to calculate a score for each iteration, i.e. set of CFDs, to select the set of CFDs $CFD_{best}$ to be applied in the repair process. The score resembles the percentage of CFDs with confidence equal to 1, i.e. CFDs not violating any tuple in the training data.

To apply rule based repair, we again use data context to represent target instances. Thus, we assume that we can apply the CFDs discovered using data context data for the target to detect violations and generate repair operations for sources based on the repair algorithm described in \cite{Cong2007} (line 16). Repair operations are based on attribute value modifications as they are sufficient to resolve CFD violations. In short, and following the notation in \cite{Cong2007}, if a tuple $t$ violates a CFD $\phi = (R: X \rightarrow Y, t_p)$,  composed of a FD plus a pattern tuple $t_p$, the algorithm either modifies the values of $t$ for the attributes matching the right-hand side of the FD, according to the pattern tuple, or modifies the values of some attributes of $t$ matching the left-hand side of the FD. In case of violations of $t$ with another tuple $t'$, different attribute modifications will be applied. The repair algorithm produces a repair that is as close as possible to the original dataset, by choosing, at each step (testing), to repair the attribute of a tuple $t$ with minimum repair cost. Such a cost model depends on a distance function, which in our case is based on the Damerau-Levenshtein metric.

\begin{algorithm}[t]
	\caption{Rule-based repair}\label{algo:repair}
	\begin{algorithmic}[1]
		\Require source schema $S$ and instances $I_S$ product schema $P$, set of relationships $R$, Set of data context schemas $D$ and instances $I_D$, lower bound $lb$, initial support $is$, step size $step$
		\Ensure Repaired source $S'$
		\Procedure{repair}{}
		\ForAll {$d \in D$}
			\State $CFD_{best} \gets \{\}$
            \State $score_{best}=lb$
            \State $support \gets is$
			\While{$score > score_{best}$}
				\State $CFD_d \gets generate\_cfd({d,support})$
				\State $CFD_{df} \gets filter(CFD_d,d)$
				\State $score \gets test(CFD_d,d)$
				\If{$score > score_{best}$}
					\State $score_{best} = score$
                    \State $CFD_{best} \gets CFD_{df}$
					\State $support=support-step$
				\EndIf
			\EndWhile
			\State ${CFD} \gets rewriteToSource(CFD,S,R))$
			\State $S' \gets gen\_test\_repair(S, CFD_{best})$
			\Return $S'$
		\EndFor
		\EndProcedure
	\end{algorithmic}
\end{algorithm}

\section{Experimental Evaluation} 
\label{sec:evaluation}

We present an experimental study of informing multiple data wrangling steps with different types of data context by evaluating the effect against the base case where data context is not used.

\begin{figure*}
\centering

\begin{subfigure}[b]{.32\linewidth}
	\centering
	\includegraphics[width=\linewidth]{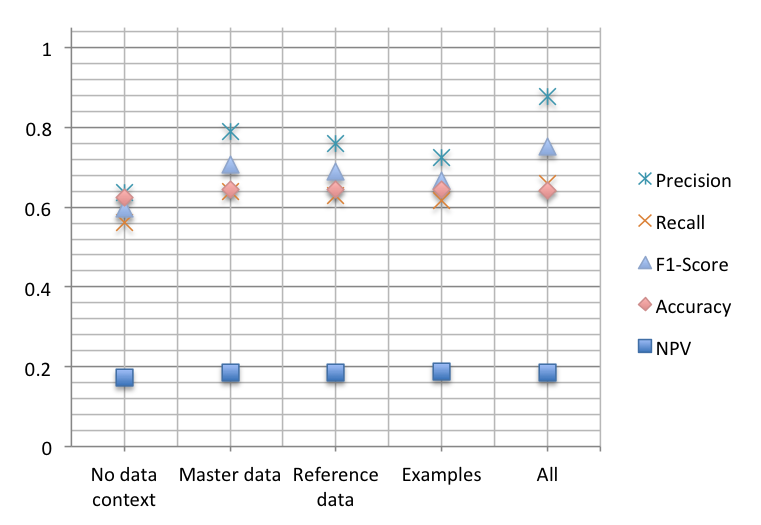}
	\caption{Data wrangling process}
	\label{fig:all}
\end{subfigure}
\begin{subfigure}[b]{.32\linewidth}
	\centering
	\includegraphics[width=\linewidth]{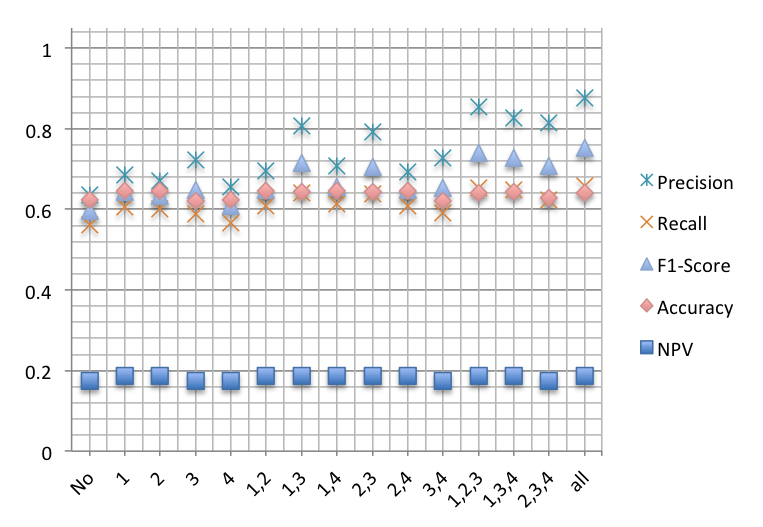}
	\caption{Wrangling stage combinations}
	\label{fig:allcomb}
\end{subfigure}
\begin{subfigure}[b]{.32\linewidth}
	\centering
    \includegraphics[width=\linewidth]{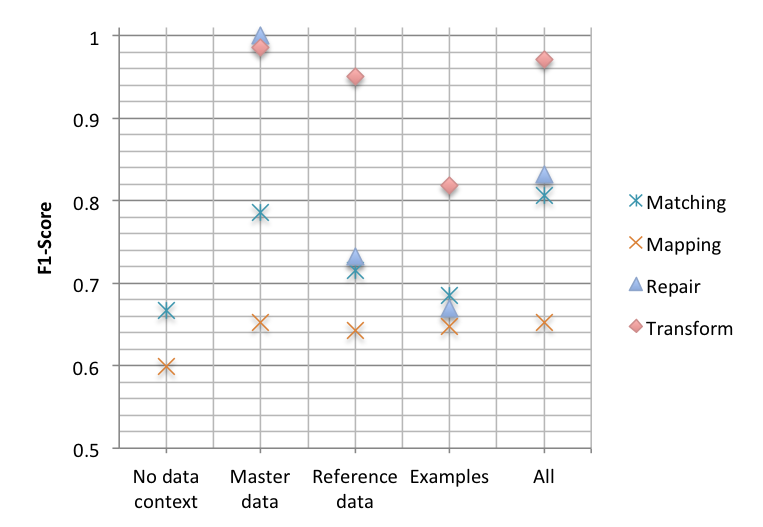}
	\caption{Individual wrangling stages}
	\label{fig:individual}
\end{subfigure}
\setlength{\belowcaptionskip}{-15pt}
\caption{Experimental results: Effect of data context types on wrangling process and on individual stages}
\end{figure*}
\setlength{\belowcaptionskip}{0pt}

\textit{Application domain and data.}
We perform our experiments on real world data consisting of web-extracted data from the real-estate domain. The data used is an extension of the running scenario described in Section \ref{sec:problem}, and consists of data from 40 real-estate agencies. All datasets have been extracted with OXPath \cite{Furche2013a} according to their representation on the web page to be as close as possible to a completely automated extraction. In addition to the real-estate data, we include a freely available deprivation statistics dataset\footnote{English deprivation indices: https://www.gov.uk/government/statistics/ english-indices-of-deprivation-2015}  for the UK. We used a subset of the extracted data with full postcode information building upon 6K tuples.

\textit{Data context.}
For reference data  we utilised the open address data set for the UK (30k tuples). It provides high quality address data including postcode, street names, city and primary and secondary address objects. For examples we used the freely available UK price paid data including 160k tuples representing property sales information. To emulate master data, we cleaned the set of extracted records from a single real-estate agency. 


\textit{Measuring wrangling quality.}
While the methods used in all stages of the wrangling process have been evaluated separately (see \cite{Aumueller2005,Marnette2011,Bogatu2017,Cong2007}), the approach of applying readily available information, data context, to them, individually or together, has not. 
To measure the quality for single stages and for the whole wrangling process we created the ground truth for the test scenario by hand. The ground truth represents the correct representation of the data the user is interested in. It provides correctly transformed, cleaned and integrated values for the set of sources according to the data context.

We use the following metrics: \textit{Precision}, the fraction of relevant items among the retrieved items:
$P = \frac{TP}{TP+FP}$; \textit{Recall}, the fraction of relevant items that have been retrieved: $R = \frac{TP}{TP+FN}$;
\textit{f-measure}, the harmonic mean of precision and recall: $F_1 = 2 * \frac{PPV * TPR}{PPV+TPR}$;
\textit{Accuracy}, the fraction of true items among all items: $ACC = \frac{TP+TN}{TP+FP+TN+FN}$;
\textit{Negative predictive value}, the fraction of negative predictions that are correct:
$NPV = \frac{TN}{TN+FN}$.

\subsection{Effect of Data Context on the Wrangling Result}
\label{subsec:eval:sequences}

The propositions to be tested are 1) that the wrangling process in total can benefit from being informed by data context, 2) that using multiple data context types together is able to improve the overall wrangling result, 3) that each data context type can be used to improve the results of at least a single step, and 4) that data context can be used to gain combined effects on multiple wrangling stages. 

We report on precision, recall, f-measure, accuracy, and NPV according to the ground truth. We compare all values of the resulting tuples with the ground truth by applying the following definitions: TP -- a value in the result that corresponds to a value in ground truth, FP -- a value in the result that does not correspond to a value in ground truth, FN -- a value in the ground truth that should be but is not in the result, TN -- a value of a source tuple that is not in the result and should not be.

We conducted experiments informing all steps of data wrangling with each data context type separately, and applying all data context types at each stage. An overview of the results is depicted in Figure \ref{fig:all}. In the case of \textit{No data context} the process executes Coma schema-based matchers, and mappings are selected at random as no instances are available to inform mapping validation. The process misses several schematic correspondences (e.g. $street$, $status$) and produces some incorrect ones ($propertyheaderbedroomandprice\_left\_h1 \equiv bedroom$). Value format transformations and rule-based repairs can't be applied without data context. By applying data context we can find additional matches, apply mapping validation, and execute format transformations and rule-based data repair. 

The experiments show that applying all data context types at once at each wrangling step results in better target values than not informing the process and than applying a single data context item. There is a gain of 0.15 in f-measure by applying all data context items at once, with precision and recall improved by 0.24 and 0.09 respectively. This substantial improvement is achieved by simply associating context data with the target schema.

The quality of the result is greatest when all data context types are available because each type holds different information to be exploited. For instance, master data and reference data items enable the process to find different schematic correspondences. Reference data enables correct matching of $street$ attributes, while master data additionally links a $status$ attribute. Reference data also supports format transformation of $street$ attributes into their desired representation. The format of the attribute $type$ can only be transformed by having examples at hand. Rule-based repair can correct values for attribute $agency$ based on master data, while reference data corrects $city$ and $street$ values. 

The results for individual data context items show that the f-measure of the target improved by 0.07, 0.09 and 0.11 (master data, reference data and examples respectively). We can report a gain in precision of 0.08, 0.12 and 0.15 and recall is increased by 0.5, 0.6 and 0.7. 
Each data context type has a positive effect and a corresponding combined effect is achieved by applying them together.

In Figure 2b we show how the target quality is affected by applying all data context items at once on different combinations of wrangling stages, e.g. schema matching and mapping generation 
(1: matching, 2: mapping, 3: transformation, 4: repair). 
The figure shows that by informing multiple steps of the process with data context, combined effects can be achieved. For instance, consider the combined effect of using data context for matching and transformation (1,3). Using data context in schema matching increases precision by 0.05, in transformations by 0.08, and in both at the same time by 0.16. Another example is that using data context for matching and repair can slightly increase the individual gains in precision from 0.05 and 0.01 to 0.07.


\subsection{Effect of Data Context on Individual Wrangling
Steps} 
\label{subsec:eval:individual}
The objective of the second experiment is to investigate how, and to what extent, each of the wrangling steps have benefited from the data context. To evaluate the effect of data context on individual stages we report on f-measure of detected matches, validated transformations and repairs, and we use the same notion of TP, FP, FN as for target quality to evaluate mappings (see Figure \ref{fig:individual}).   

By applying all data context types at once, a gain of 0.13 in f-measure can be achieved as the detected matches are partly cumulative. 
Most of the additional matches can be detected by domain recognisers. For instance, using master data, additional matches for the target attribute price are detected. In general, domain recognisers work well to detect matches for address data including streets and cities.

Utilising data context for mapping validation leads to an increase of 0.05 to 0.06 in f-measure. If multiple data context types are available, we select the
mapping with the highest target coverage and the highest verification score based on the data context instances. 
The reported f-measure on validated transformations when applying all data context types at once is 0.97, for reference and master data we achieved 0.95 and 0.98, while examples lead to 0.81. It appears that the combination of sources and examples is, in some cases, not expressive enough to enable the required transformation to be synthesised. 
The f-measure in data repair depends on the data context (master data: 1, reference data: 0.73, examples: 0.66, all: 0.83) applied. The lower f-measure for reference data and examples occurs because some opportunities for repairs are missed. 

\section{Conclusions} 
\label{sec:conclusion}

Data scientists have been found to be spending as much as 80\% of their time on data wrangling\footnote{New York Times://http://nyti.ms/1Aqif2X}, so cost-effective data wrangling is crucial to the successful use of big data. In this paper we have shown in a representative real world example an improvement in precision of 24\%, combined with a 9\% increase in recall, by extending a four step wrangling process to use data context throughout. As illustrated in our demo paper, data scientists can associate data context with a target schema with modest effort~\cite{Konstantinou2017}. As such this paper has presented a methodology for enhanced automation that provides a significant return on investment.

Specifically, readily available types of contextual information, such as master data, reference data, and examples, have been used to improve the outcome of all steps in our data wrangling process in different ways: in {\it matching} by extending the collection of matchers that can be applied; in {\it mapping} by allowing mapping validation to be informed by the results of verification; and in both {\it format transformation} and {\it repair} by enabling rules to be learned. We showed that applying multiple data context types on several wrangling stages results in a combined gain in the quality of the final wrangling result; firstly, by adding up the effect of different data context types within a stage, and secondly, by accumulating results from different stages. 

\noindent
\textit{Acknowledgment}
We thank the UK EPSRC for their support through the VADA Programme Grant.

\def\bibfont{\footnotesize}
\footnotesize
\bibliographystyle{IEEEtran}
\bibliography{ms}

\end{document}